\def\OA{{(\Omega, A)}}

\documentclass[aps,prl,twocolumn,groupedaddress,showkeys,showpacs,superscriptaddress,floatfix]{revtex4-2}

\usepackage{amsmath}
\usepackage{graphicx}
\usepackage{xcolor}

\begin{document}


\title{Generation of travelling sine-Gordon breathers in noisy long Josephson junctions}



\author{Duilio De Santis}
\email[]{duilio.desantis@unipa.it}
\affiliation{Department of Physics and Chemistry ``E.~Segr\`{e}", Group of Interdisciplinary Theoretical Physics, University of Palermo, I-90128 Palermo, Italy}

\author{Claudio Guarcello} 
\affiliation{Department of Physics ``E.~R.~Caianiello", University of Salerno, I-84084 Fisciano, Salerno, Italy}

\author{Bernardo Spagnolo}
\affiliation{Department of Physics and Chemistry ``E.~Segr\`{e}", Group of Interdisciplinary Theoretical Physics, University of Palermo, I-90128 Palermo, Italy}
\affiliation{Radiophysics Department, Lobachevsky State University, 603950 Nizhniy Novgorod, Russia}

\author{Angelo Carollo}
\affiliation{Department of Physics and Chemistry ``E.~Segr\`{e}", Group of Interdisciplinary Theoretical Physics, University of Palermo, I-90128 Palermo, Italy}

\author{Davide Valenti}
\affiliation{Department of Physics and Chemistry ``E.~Segr\`{e}", Group of Interdisciplinary Theoretical Physics, University of Palermo, I-90128 Palermo, Italy}


\date{\today}

\begin{abstract}

The generation of travelling sine-Gordon breathers is achieved through the nonlinear supratransmission effect in a magnetically driven long Josephson junction, in the presence of losses, a current bias, and a thermal noise source. We demonstrate how to exclusively induce breather modes by means of controlled magnetic pulses. A nonmonotonic behavior of the breather-only generation probability is observed as a function of the noise intensity. An experimental protocol providing evidence of the Josephson breather's existence is proposed.

\end{abstract}


\maketitle


\emph{Introduction.}{\textemdash}The sine-Gordon~(SG) equation is one of the most celebrated nonlinear wave models, with a domain of application that includes pendula, Josephson junctions~(JJs), gravity, high-energy physics, biophysics, and seismology~\cite{Jesus_2014, Ivancevic_2013, Yakushevich_2021, Bykov_2014}. Interestingly, solitonic excitations are sustained in a SG medium, \emph{kinks} and \emph{breathers} being the basic building blocks. Kinks, i.e., topological solitons, stem from an inherent degeneracy of the system's ground state, while breathers are space-localized, time-periodic modes that can result from the attractive interaction between kinks and antikinks~\cite{Scott_2003, Dauxois_2006}. Indeed, the SG equation provides an ideal environment for the exploration of soliton dynamics, with subsequent striking experimental realizations, thus it continues to attract great research interest in many scientific areas.

The electrodynamics of a long Josephson junction~(LJJ) can be accurately described in the SG framework. Here, a kink (or fluxon) represents a quantum of magnetic flux~${ \Phi_0 }$. Since it can be stored, steered, manipulated, and made to interact with electronic instrumentation, such a nonlinear wave is featured in many applications~\cite{Ustinov_1998, Guarcello_2017, Gua18, GuaSolBra18, Wustmann_2020, Osborn_2021}. Notably, the quantum dynamics of a single fluxon has been experimentally demonstrated~\cite{Wallraff_2003}.

Breathers are far more elusive than kinks, mainly because they radiatively decay due to dissipation, unless particular forms of driving are employed, and because their oscillatory nature gives rise to a practically null average voltage, i.e., beyond sensitivity of the existing high-frequency oscilloscopes~\cite{Gulevich_2012}. As a consequence, despite the large variety of numerical and theoretical studies devoted to them~\cite{Kivshar_1989, Gulevich_2006, Johnson_2013}, experimental evidence of breather modes has yet to be found in LJJs.

The interest towards breathers is also motivated by their significant applicative potential. In fact, such a mode could be effectively used to develop some novel applications in information transmission~\cite{Macias-Diaz_2007}. Furthermore, in contrast to kinks, they possess an internal degree of freedom, i.e., a proper frequency, which is particularly valuable for quantum computation purposes. More specifically, breathers behave as macroscopic artificial two-level atoms in an LJJ with a small capacitance per unit length, so that the realization of a Josephson breather qubit has been proposed~\cite{Fujii_2007, Fujii_2008}. To these ends, detailed investigations concerning possible generation and control mechanisms are crucial.

The excitation and detection of breather-like objects has been deeply studied in the context of discrete systems as well. In particular, in JJ parallel arrays, which are modelled by the discrete SG equation (also known as the Frenkel–Kontorova model), the existence of oscillobreathers has been predicted, but due to their rapid pulsations, an experimental confirmation is still missing~\cite{Mazo_2003, Jesus_2014}. Instead, the so-called rotobreather states can be tracked by measuring the local dc voltages throughout the JJ array, and they have been successfully observed in JJ ladders~\cite{Trias_2000, Binder_2000}.

This letter proposes the use of magnetic pulses for the generation of travelling breather modes into an LJJ by means of the nonlinear supratransmission~(ST) effect. According to the latter phenomenon, a nonlinear system subjected to a sinusoidal driving with frequency laying in the forbidden band gap~(FBG) can support energy transmission in the form of solitonic excitations, if the forcing amplitude is strong enough. Initially discussed in a discrete SG chain~\cite{Geniet_2002, Geniet_2003}, the ST mechanism appears to be the result of a generic nonlinear instability~\cite{Leon_2003}, and the corresponding activation threshold has been found in various situations~\cite{Khomeriki_2004_1, Khomeriki_2004_2, Bodo_2009, Vasilescu_2010}.

Here, by looking at the average voltage drop across the junction, the external pulse's frequency/amplitude space is thoroughly analyzed to exploit the ST process as a controllable source of breathers in LJJs. Interestingly, in the presence of dissipation, a current bias, and a thermal noise source, vast regions associated with the exclusive emergence of breather excitations are observed. In particular, the study reveals a sort of fluctuation-induced widening of these areas, which is seen to occur in correspondence with a nonmonotonic behavior of the breather-only generation probability, defined below, as a function of the noise amplitude. The latter result highlights the effectiveness of noise as a control parameter in the LJJ device.

Based on these findings, an experimental procedure providing evidence of the existence of breathers in LJJs is outlined. Moreover, given the degree of universality of the ST process, and its robustness against discreteness and finiteness, the application of the proposed approach in a discrete domain naturally comes to mind. Among the many contexts in which energy localization and transmission is being actively investigated~\cite{Flach_2008, Dmitriev_2016, Liu_2021}, one could look, e.g., at oscillobreathers in a JJ parallel array.

\emph{The model.}{\textemdash}An overlap-geometry Josephson tunnel junction is considered, assuming that the dynamics of the phase difference between the pair wave functions of the two superconductors ${ \varphi (x, t) }$ follows the equation~\cite{Barone_1982, Castellano_1996}
\begin{equation}
\label{eqn:1}
\varphi_{xx} - \varphi_{tt} - \alpha \varphi_{t} = \sin \varphi - \gamma - \gamma_T (x, t) .
\end{equation}
The ${ x = 0 }$ end of the junction is subjected to an oscillating magnetic field, perpendicular to the length of the junction and parallel to the plane of the barrier, thus the following boundary conditions apply
\begin{equation}
\label{eqn:2}
\varphi_x (0, t) = \widetilde{A} (t) \sin \left( \Omega t \right) , \; \; \; \varphi_x (l, t) = 0 .
\end{equation}
A subscript notation is used to denote partial differentiation, space is normalized to the Josephson penetration depth, ${ \lambda_J = \sqrt{ \Phi_0 / \left( 2 \pi J_c L_P \right)} }$, and time is normalized to the inverse of the Josephson plasma frequency, ${ \omega_p = \sqrt{ 2 \pi J_c / \left( \Phi_0 C \right) } }$, where $ J_c $ is the critical value of the Josephson current density, ${ L_P }$ is the inductance per unit length, and ${ C }$ is the capacitance per unit length. In Eq.~\eqref{eqn:1}, ${ \alpha = G / \left( \omega_p C \right) }$ is a damping parameter, $ G $ is an effective normal conductance, ${ \gamma = J_b / J_c }$ is the normalized bias current, and ${ \gamma_{T} (x, t) }$ is a Gaussian, zero-average noise source with the autocorrelation function given by
\begin{equation}
\label{eqn:3}
\langle \gamma_{T}(x_1, t_1) \gamma_{T}(x_2, t_2) \rangle = 2 \alpha \Gamma \delta (x_1 - x_2) \delta (t_1 - t_2) .
\end{equation}
Here, ${ \Gamma = 2 e k_B T / \left( \hbar J_c \lambda_J \right) }$ is the noise amplitude, which is proportional to absolute temperature $ T $ ($ e $ is the electron charge, $ k_B $ is the Boltzmann constant, and $ \hbar $ is the reduced Planck constant). Finally, in Eq.~\eqref{eqn:2}, ${ \widetilde{A} (t) }$ and ${ \Omega }$ are two dimensionless quantities related, respectively, to the external field's amplitude and frequency, and ${ l = L / \lambda_J }$ is the normalized length of the junction.

Equation~\eqref{eqn:1} is numerically solved with the initial conditions
\begin{equation}
\label{eqn:4}
\varphi (x, 0) = \arcsin \gamma , \; \; \; \varphi_t (x, 0) = 0 ;
\end{equation}
also, in order to access the ST phenomenon, the external field has to oscillate at a frequency falling into the junction's FBG, i.e., ${ \Omega < 1 }$, corresponding to ${ \omega < \omega_p \sim 100 \; \textrm{GHz} - 1 \; \textrm{THz} }$~\cite{Gulevich_2006, Gulevich_2012}. Moreover, to reproduce a meaningful experimental pulse, Gaussian \textit{switching-on/off regimes} are chosen as
\begin{equation}
\label{eqn:5}
\widetilde{A} (t) = \begin{cases}
	A \exp \left[ - \frac{(t - t_{\rm{on}})^2}{2 \sigma_{\rm{on}}^2} \vphantom{\exp \left( - \frac{(t - t_{\rm{off}})^2}{2 \sigma_{\rm{off}}^2} \right)} \right] & t < t_{\rm{on}} \\
	A & t_{\rm{on}} \leq t < t_{\rm{off}} \\
 A \exp \left[ - \frac{(t - t_{\rm{off}})^2}{2 \sigma_{\rm{off}}^2} \right] & t \geq t_{\rm{off}} .
 \end{cases}
\end{equation}
In Eq.~\eqref{eqn:5}, the Gaussian distribution with standard deviation $ \sigma_{\rm{on}} $ provides a smoothly increasing signal envelope for ${ t < t_{\rm{on}} }$ (in practice, ${ t_{\rm{on}} = 3 \sigma_{\rm{on}} }$ is set). Then, the boundary of the junction is sinusoidally driven until an induced travelling excitation reaches a selected position~\footnote{Since the medium's FBG is being considered, one may detect a nonlinear mode's presence by looking at the modulus of the phase ${ \left\lvert \varphi \right\rvert }$ at the designated position, which should be taken relatively close to ${ x = 0 }$.}, i.e., for ${ t < t_{\rm{off}} }$. If this event occurs, the driving amplitude is gradually decreased, with the typical scale $ \sigma_{\rm{off}} $. By doing so, one is mostly able to achieve the controlled generation of single breather modes in the system.

\begin{figure}[t!!]
\includegraphics[width=0.75\columnwidth]{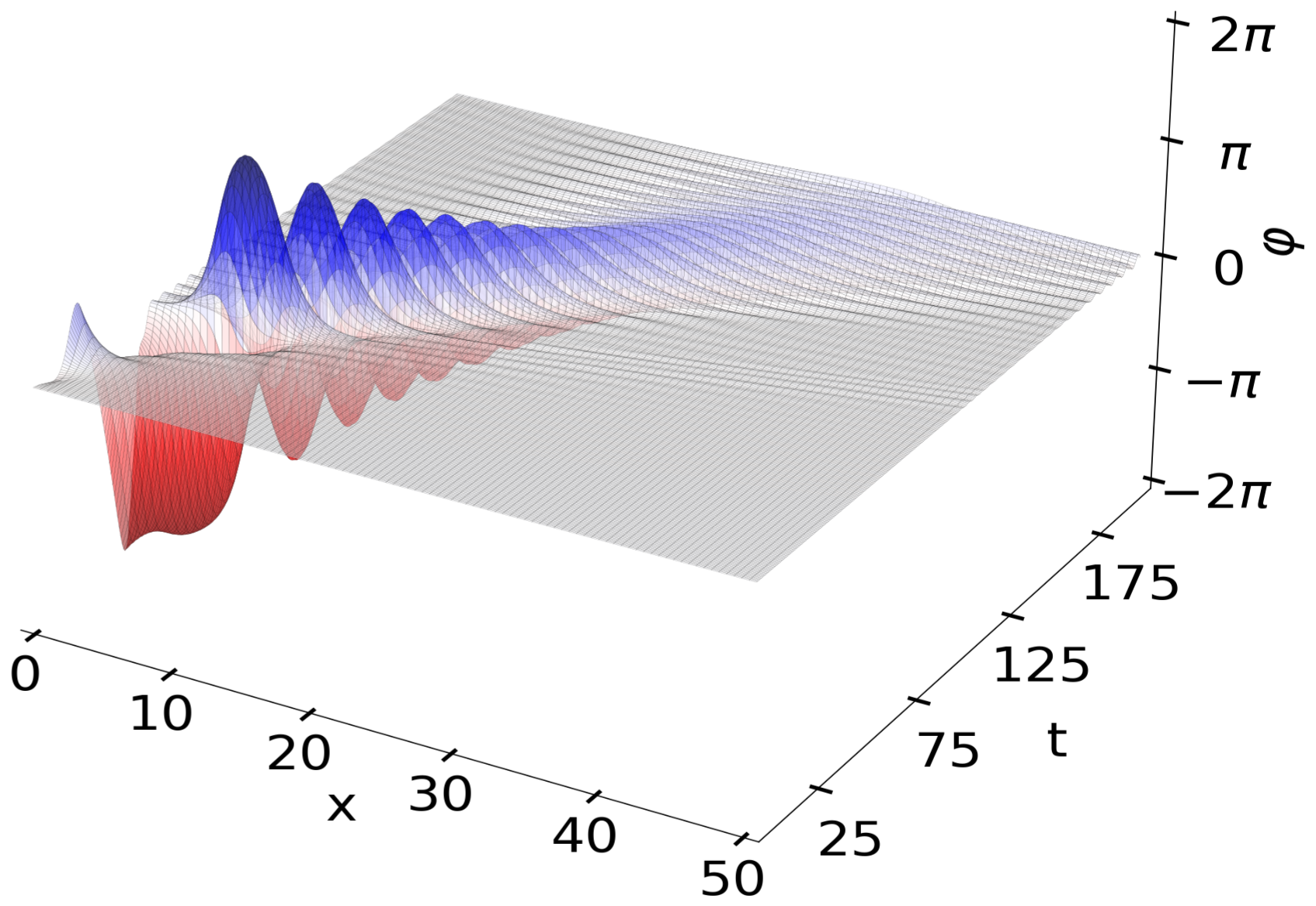}
\caption{Time evolution of an ST-induced travelling breather in the presence of dissipation. Parameter values: ${ t_{\rm{max}} = 200 }$ (observation time), ${ \Omega = 0.25 }$, and ${ A = 2.09 }$.}
\label{fig:0}
\end{figure}

In what follows, the length of the junction is ${ l = 100 }$, which generally allows to ignore reflection effects at ${ x = l }$, while keeping reasonable execution times. Also, the damping coefficient is ${ \alpha = 0.02 }$~\cite{Lomdahl_1982}, the bias current is ${ \gamma = 0 }$, and the width of the increasing (decreasing) Gaussian tail is ${ \sigma_{\rm{on}} = 10 }$ (${ \sigma_{\rm{off}} = 2.5 }$). A typical simulation outcome is illustrated in Fig.~\ref{fig:0}.

\emph{Breather detection.}{\textemdash}As stated above, one of the main aims of the present work is to identify the regions of the $ \OA $ parameter space in which the ST generation mechanism leads to breather excitations only. To this end, the meaning of the expression ``breather-only'' used throughout the letter shall now be operatively defined. Essentially, such a term includes all the cases in which breathers are the only solitons left in the junction from some point onwards, after a possible transitory phase that could involve kinks and antikinks. Conversely, the following situations are not considered to belong to the breather-only class: \emph{i})~at least one bound kink-antikink couple gets separated by the end of the simulation; \emph{ii})~at least one unpaired kink (or antikink) is produced; \emph{iii})~the breather state is not observed at all.

Different techniques can be constructed for the classification of the outcomes of the simulations. For instance, one can take advantage of the practically null average voltage drop across the junction produced by breathers. By virtue of the second Josephson relation~\cite{Barone_1982}, it is possible to define the (normalized) time-averaged voltage
\begin{equation}
\label{eqn:6}
\left\langle V \right\rangle = \Big\langle \int_0^l \varphi_t dx \Big\rangle .
\end{equation}
Through extensive preliminary tests (the values of the simulation parameters are given above) it was seen that, in the numerical runs where some excitation is induced, the exclusive emergence of breather modes takes place if the condition
\begin{equation}
\label{eqn:7}
\left\lvert \langle V \rangle \right\rvert \lesssim 0.05
\end{equation}
is verified~\footnote{In practice, due to possible reflection effects at the boundaries of the system, the values ${ \left\lvert \varphi (0, t) \right\rvert }$ and ${ \left\lvert \varphi (l, t) \right\rvert }$ require additional monitoring.}, otherwise at least one kink (or antikink) is left in the junction at the end of the simulation, i.e., at ${ t = t_{\rm{max}} }$. Equation~\eqref{eqn:7} will be indicated hereafter as the \textit{voltage criterion}.

\emph{Results and discussion.}{\textemdash}To numerically integrate Eq.~\eqref{eqn:1}, a Fortran implementation of the implicit finite-difference scheme detailed in Ref.~\cite{Lomdahl_1982} is used, with the noise term approximated according to Ref.~\cite{Tuckwell_2016}. The discretization steps are ${ \Delta x = \Delta t = 0.005 }$, and the observation time is usually set to ${ t_{\rm{max}} = 200 }$ in order to follow the entire evolution of the emitted breathers.

\begin{figure}[t!!]
\includegraphics[width=0.75\columnwidth]{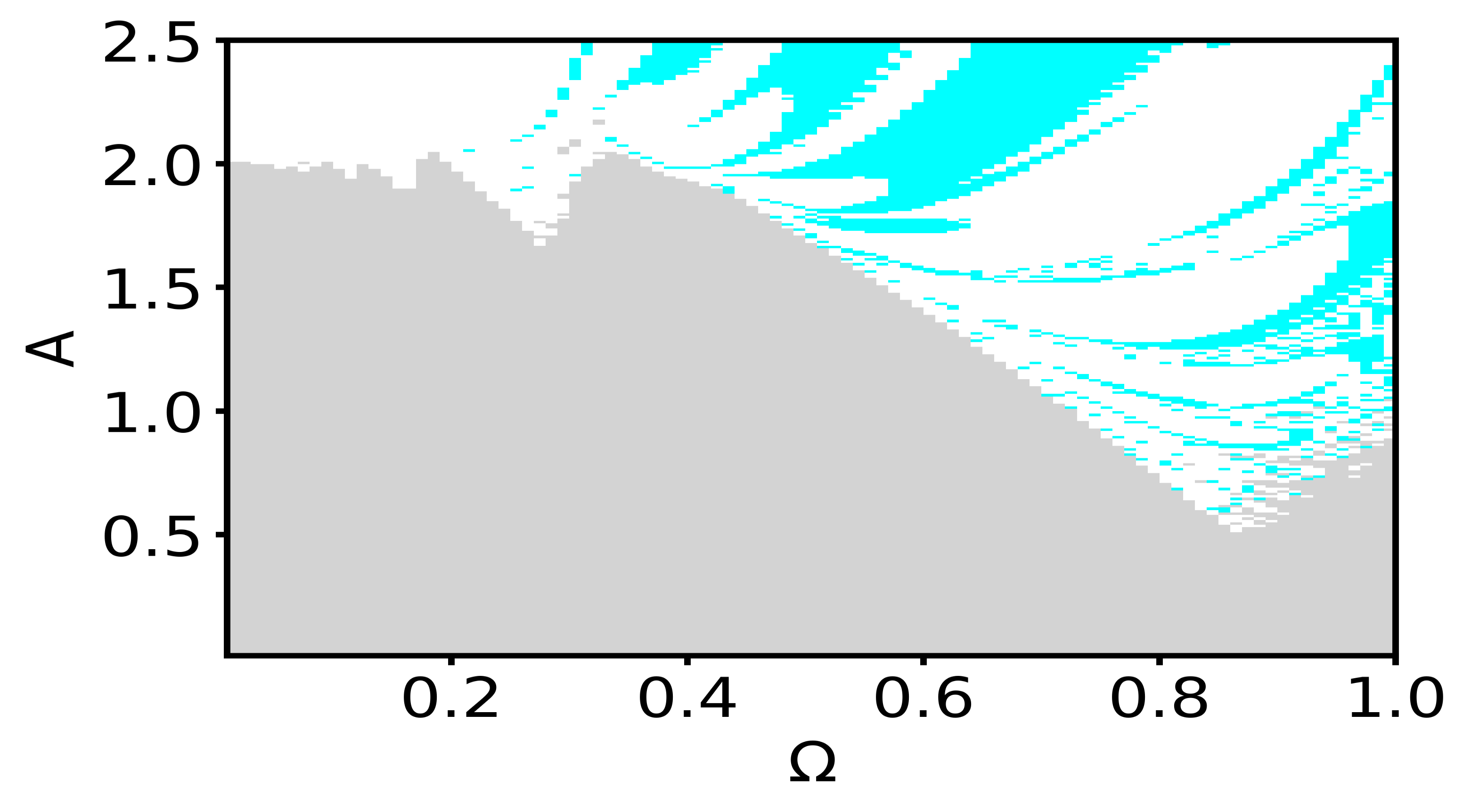}
\caption{Bifurcation diagram in the $ \OA $ plane. The gray color indicates $ \OA $ couples for which no excitations of clear solitonic nature are detected in the medium, the white color corresponds to regions with at least one kink (or antikink) left in the junction at ${ t = t_{\rm{max}} = 200 }$, and the cyan color is exclusively associated with breathers according to the voltage criterion [Eq.~\eqref{eqn:7}].}
\label{fig:1}
\end{figure}
For the first set of simulations, the stochastic term ${ \gamma_T (x, t) }$ is not considered. In this case, the $ \OA $ parameter space is explored by varying the driving frequency $ \Omega $ in ${ \left[ 0, 1 \right] }$ and the amplitude $ A $ in ${ \left[ 0, 2.5 \right] }$, with increments ${ \Delta \Omega = \Delta A = 0.01 }$. Figure~\ref{fig:1} presents the results obtained through the voltage criterion [Eq.~\eqref{eqn:7}] in the form of a refined bifurcation diagram. Along with the essential features of the ST effect, the plot clearly displays the presence of significant breather-only (cyan) areas, while the gray color indicates $ \OA $ couples for which no excitations of clear solitonic nature are detected in the medium, and the white color corresponds to regions with at least one kink (or antikink) left in the junction at ${ t = t_{\rm{max}} }$.

Upon changing the values of the parameters ${ \alpha }$, ${ \gamma }$, ${ \sigma_{\rm{on}} }$, and ${ \sigma_{\rm{off}} }$, the ranges in which the overall picture is not substantially altered can be found (not shown here). Naturally, as the bias current or the duration of the switching-off regime are increased, the breather-only regions are seen to gradually disappear; in fact, the ${ \gamma }$ perturbation term tends to dissociate a breather into a kink-antikink pair~\cite{Gulevich_2012}, and additional kinks (and/or antikinks) are more likely generated if the forcing signal takes a longer time to stop injecting energy into the medium, i.e., for higher values of ${ \sigma_{\rm{off}} }$. Indicatively, one would choose ${ \gamma \lesssim 0.1 }$ and ${ \sigma_{\rm{off}} \lesssim 10 }$. The shape and localization of the zones of interest may also slightly change if the parameters ${ \alpha }$ and ${ \sigma_{\rm{on}} }$ are varied, but the main features of Fig.~\ref{fig:1} remain. Moreover, further testing indicates that qualitatively analogous breather-only areas can be obtained with fixed-duration forcing envelopes as well, i.e., with pulses that do not automatically shut down after an externally-induced excitation is detected.

\begin{figure}[t!!]
\includegraphics[width=\columnwidth]{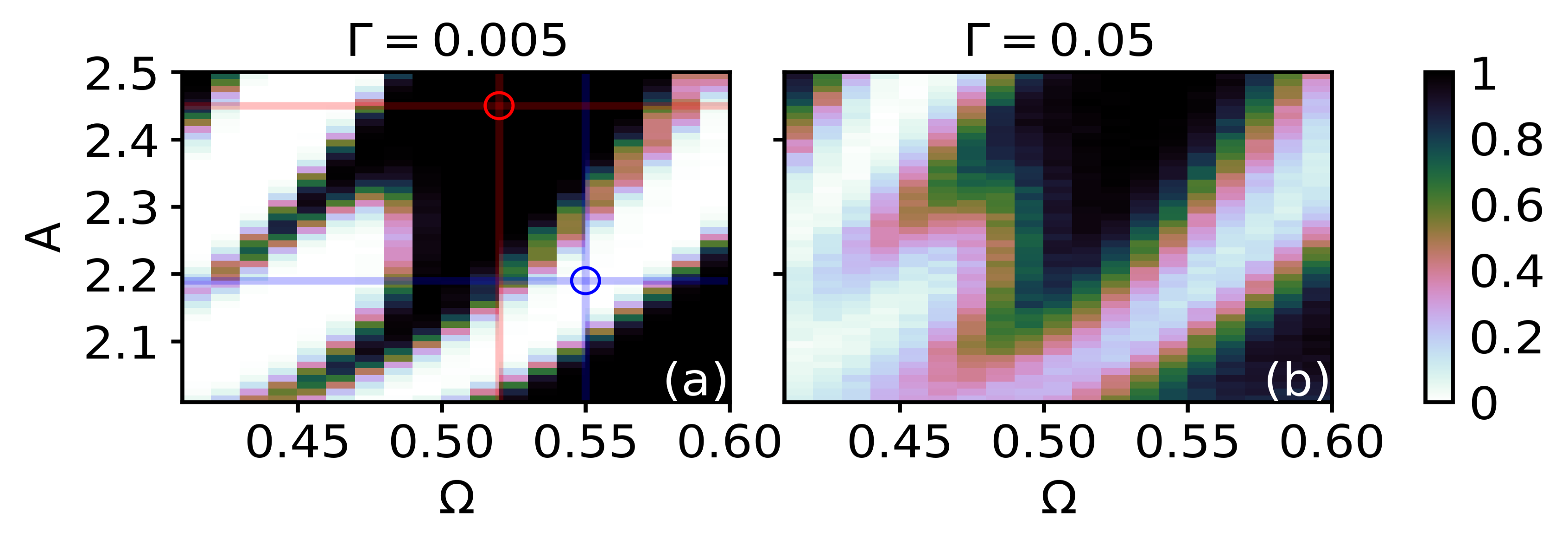}
\caption{Breather-only generation probability estimated over ${ N = 512 }$ realizations according to the voltage criterion [Eq.~\eqref{eqn:7}]. Here, ${ t_{\rm{max}} = 150 }$. Moreover, ${ \Gamma = 0.005 }$ in panel~(a) and ${ \Gamma = 0.05 }$ in panel~(b). The blue (red) circle in panel~(a) identifies the combination ${ \Omega = 0.55 }$ and ${ A = 2.19 }$ (${ \Omega = 0.52 }$ and ${ A = 2.45 }$).}
\label{fig:2}
\end{figure}
A crucial aspect to discuss is whether the breather-only regions are robust enough to survive in a noisy environment. Denoting the number of breather-only cases identified by means of the voltage criterion [Eq.~\eqref{eqn:7}] with ${ N_{\rm{breather-only}} }$ and the total number of realizations performed with $ N $, a breather-only generation probability is defined here as ${ P_{\rm{breather-only}} = \lim_{N \to \infty} N_{\rm{breather-only}} / N }$. Estimations of such a quantity over ${ N = 512 }$ realizations are presented in Fig.~\ref{fig:2} for the noise amplitudes ${ \Gamma = 0.005 }$ and ${ 0.05 }$~\cite{Guarcello_2017}, with ${ \Omega \in \left[ 0.4 , 0.6 \right] }$, ${ A \in \left[ 2 , 2.5 \right] }$, and ${ \Delta \Omega = \Delta A = 0.01 }$. The latter restriction, taken due to the extent of the computational task, fully contains a single deterministic breather-only zone, along with the edges of the confining ones, see Fig.~\ref{fig:1}.

A first remark concerning the panels of Fig.~\ref{fig:2} is that the breather-only regions are still clearly recognizable, since the generation probability quickly rises in their vicinity, see the black areas. As the value of $ \Gamma $ is increased, a progressively smaller core of high probability is expected to survive because of the larger current fluctuations that tend to break up the breather state, and this is indeed observed. Nevertheless, a sort of noise-induced widening of these structures is seen as well, because most of the neighboring $ \OA $ combinations, that deterministically are not associated with the formation of breather modes only, acquire nonzero probability values. Eventually, for ${ \Gamma \sim 1 }$, different kinds of excitations (including breathers, kinks, and antikinks) begin to appear in the junction, in addition to the ST-induced ones, due to the magnitude of the stochastic perturbation, and the breather-only generation probability falls to zero in the $ \OA $ parameter space in a uniform way.

\begin{figure}[t!!]
\includegraphics[width=\columnwidth]{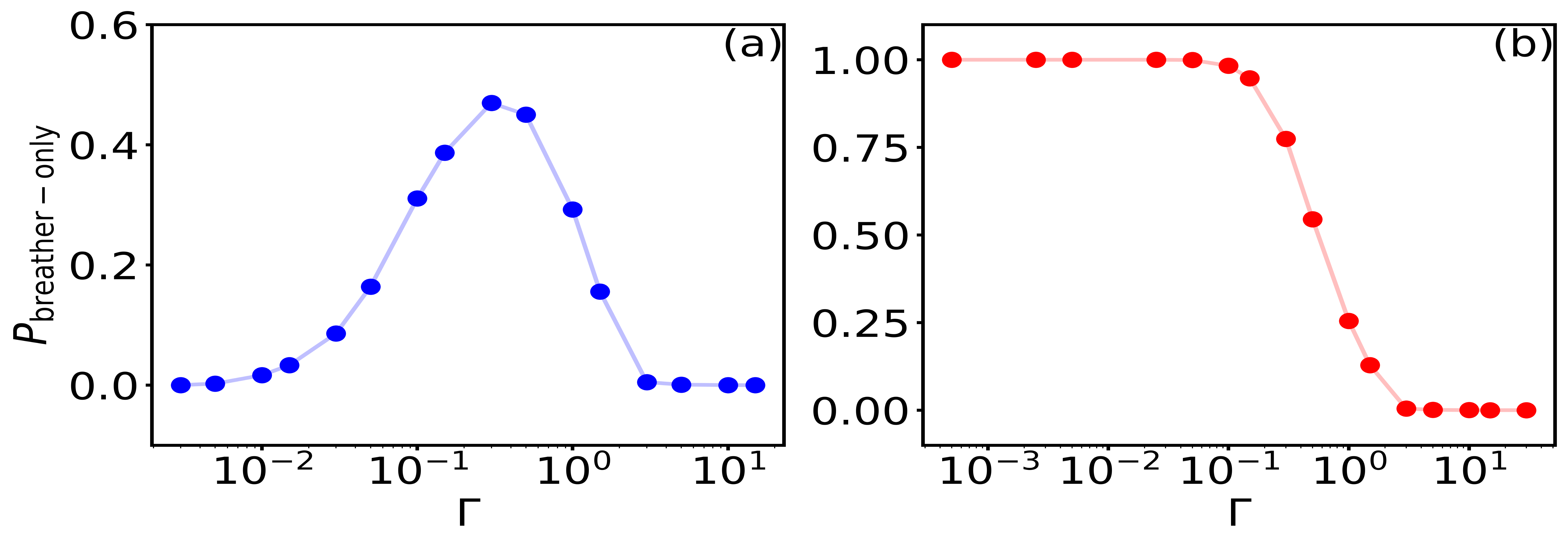}
\caption{Breather-only generation probability estimated over $ N = 10^4 $ realizations, according to the voltage criterion [Eq.~\eqref{eqn:7}], as a function of the noise amplitude $ \Gamma $. Here, ${ t_{\rm{max}} = 200 }$. Moreover, the combination ${ \Omega = 0.55 }$ and ${ A = 2.19 }$ is chosen in panel~(a), whereas ${ \Omega = 0.52 }$ and ${ A = 2.45 }$ in panel~(b).}
\label{fig:3}
\end{figure}
To further illustrate its behavior, the quantity ${ P_{\rm{breather-only}} ( \Gamma ) }$ is calculated for ${ N = 10^4 }$ realizations and the two sample $ \OA $ combinations circled in Fig.~\ref{fig:2}(a). In the first case (${ \Omega = 0.55 }$ and ${ A = 2.19 }$), which falls outside of the breather-only region identified for $ \Gamma = 0 $, a nonmonotonic profile is obtained; as shown in Fig.~\ref{fig:3}(a), the probability is practically zero when ${ \Gamma \to 0 }$ and the dynamics is closer to the deterministic case, it reaches a peak for an optimal value of $ \Gamma $, and eventually goes back to zero when the stochastic influence becomes disruptive, i.e., when the fluctuations are sufficiently strong to both easily break the oscillatory bound state and produce additional excitations into the junction. The second combination (${ \Omega = 0.52 }$ and ${ A = 2.45 }$), taken as a representative case for the high-probability core, leads to a decreasing steplike function, see Fig.~\ref{fig:3}(b). This determines the temperature values at which breather modes can be safely excited without noise disturbances.

In view of possible applications, the robustness of the induced breathers against the stochastic background is also quite relevant. Although not explicitly shown here, a detailed analysis indicates that there exists a wide range of noise intensities where their average persistence time above the level of fluctuations remains close to the quantity ${ 1 / \alpha }$. The latter is the perturbative prediction on the breather's radiative decay lifetime for ${ \Gamma = 0 }$~\cite{Kivshar_1989, Gulevich_2006}, which is seen to hold independently of the position in the $ \OA $ plane.

Now, based on these results, a test giving experimental evidence of breather modes in LJJs is outlined. First, scanning a portion of the $ \OA $ space in the presence of a sufficiently high current bias allows for the characterization of the ST process, if suitable average voltage measurements can be performed. In fact, the power balance between dissipation and input from the current bias term leads to well-defined voltage patterns whenever fluxons emerge (even from a broken up breather state), while very low average voltages are expected for the $ \OA $ couples that cannot transmit energy in the medium. Once established that nonlinear (fluxon-based) excitations are indeed being induced above a certain $ \OA $ threshold, such a region can be inspected for breather-only $ \OA $ pairs, whose existence and robustness is demonstrated earlier in the letter. Specifically, in the unbiased junction a potential breather-only $ \OA $ combination produces a very low average voltage. If the latter case is encountered, a sufficiently high current bias ($ \gamma \gtrsim 0.1 $, see, e.g., Ref.~\cite{Gulevich_2012}) should be switched on after a time interval much greater than ${ 1 / \alpha }$ from the magnetic pulse's application. If the low average voltage persists, breathers only are quite likely being generated, and due to their radiative decay, no current-driven dissociation into the kink-antikink state can take place. Conversely, if fluxon-related voltage patterns arise after the current switch, indicating the presence of at least one kink (or antikink), another $ \OA $ couple can be tried. To summarize, in the case of a persisting low average voltage, one is observing high-frequency excitations which vanish within a time interval ${ \sim 1 / \alpha }$. Linear~(plasma) waves can be reasonably excluded since the junction's FBG is being considered, and besides the preliminary $ \OA $ sweep confirms that fluxon-based modes are being excited. The existence of Josephson breathers would be supported by this observation.

\emph{Conclusions.}{\textemdash}The present work shows that magnetic pulses can be a controllable source of travelling breather modes in an LJJ. More precisely, by looking at the average voltage drop across the junction, refined bifurcation diagrams are produced in the driving pulse's $ \OA $ space, in the presence of dissipation, a current bias, and a thermal noise source. This allows for the characterization of significant regions where the exclusive formation of breathers occurs. Moreover, these areas are seen to maintain their identity in the noisy case, and a sort of noise-induced widening is found as well. The latter fact happens in correspondence with a nonmonotonic behavior of the breather-only generation probability, defined above, as a function of the noise amplitude.

Studies that focus on generation and control techniques are vital for all breather-related applications, e.g., in information transmission~\cite{Macias-Diaz_2007} and quantum computation~\cite{Fujii_2007, Fujii_2008}. Furthermore, given that the detection of Josephson breathers is a long-standing problem in the realm of mesoscopic soliton physics, an experimental strategy to confirm their existence is proposed.

Finally, the degree of universality of the ST process, together with its robustness against discreteness and finiteness, suggests applying this general approach to the discrete world. For example, interesting developments can be expected for oscillobreathers in JJ parallel arrays, but energy localization and transmission is being actively investigated in many other contexts as of today~\cite{Flach_2008, Dmitriev_2016, Liu_2021}.

\end{document}